\documentclass[10pt]{elsarticle}
\usepackage[margin=3.2cm]{geometry}
\usepackage{hyperref}
\usepackage{amsmath}
\usepackage{xcolor}
\usepackage{newtxtext}
\usepackage{newtxmath}

\biboptions{authoryear}

\begin{document}

\begin{frontmatter}

\title{Heavy and light inertial particle aggregates in homogeneous isotropic turbulence: A study on breakup and stress statistics}

\author[inst1]{Graziano Frungieri\corref{gf}}
{\corref{mycorrespondingauthor}\fnref{gf}}
\cortext[mycorrespondingauthor]{Corresponding author}
\ead{graziano.frungieri@tum.de}
\fntext[gf]{graziano.frungieri@tum.de, Tel: +49 8161 71-3788}
\author[inst2]{Matth\"aus U. B\"abler}
\author[inst3]{Luca Biferale}
\author[inst4]{Alessandra Sabina Lanotte}
\address[inst1]{Chair of Process Systems Engineering, TUM School of Life Sciences, Technical University of Munich}
\address[inst2]{Department of Chemical Engineering, KTH Royal Institute of Technology, Stockholm, Sweden}
\address[inst3]{Department of Physics and INFN, University of Tor Vergata, Rome, Italy}
\address[inst4]{CNR NANOTEC and INFN, Sez. Lecce, Lecce, Italy}

\begin{abstract}
{The breakup of inertial, solid aggregates in an incompressible,
  homogeneous and isotropic three-dimensional turbulent flow is
  studied by means of a direct numerical simulation, and by a
  Lagrangian tracking of the aggregates at varying Stokes number and
  fluid-to-particle density ratio.  Within the point-particle
  approximation of the Maxey-Riley-Gatignol equations of motion, we
  analyse the statistics of the time series of shear and drag
  stresses, which are here both deemed as responsible for particle
  breakup.  We observe that, regardless of the Stokes number, the
  shear stresses produced by the turbulent velocity gradients
  similarly impact the breakup statistics of inertial and neutrally
  buoyant aggregates, and dictate the breakup rate of loose
  aggregates. When the density ratio is different from unity, drag
  stresses become dominant and are seen to be able to cause to breakup
  of also the most resistant aggregates. The present work paves the
  way for including the role of inertia in population balance models
  addressing particle breakup in turbulent flows.}
\end{abstract}


\end{frontmatter}

\section{Introduction}
Breakup of particles dispersed in a fluid flow is found at the core of
many natural and engineering processes.  For instance, in aquatic
systems, the breakup of plastic waste governs the rate of
microparticle production and, as such, plays a key role on the
microplastics rate of release in the ocean \citep{garvey2020molecular,
  poulain2018small,Villermaux2021}.  In some pharmaceutical
applications, active particles are in need to be reduced in size
before administration can take place
\citep{capecelatro2022recent,sabia2022novel,vasquez2022response}, and
in polymer compounding processes, controlled breakup and
redistribution of filler agglomerates is used to produce composites
with enhanced mechanical and/or thermal properties
\citep{frungieri2020CFDDEM, frungieri2022cfd}.

The breakup of dispersed particles is determined by a number of
phenomena that challenges simple modelling approaches
\citep{baebler2008modelling}. The origin of this complexity relies on
few main features: the first is the multi-scale nature of the problem,
with relevant spatial scales ranging from the micron size of the
particles to possibly hundreds of meters (the integral scale of the
flow); the second is associated to the way aggregates spatially sample
the flow, and the third depends on the complex interplay between
fluid-induced stresses and inter-particle cohesive forces, which
eventually governs the aggregate breakup dynamics.

Depending on the application of interest and degree of insight needed,
different methods can be deemed as suitable to model breakup, and they
can be mainly differentiated on the basis of the treatment of the
dispersed and dispersing phase and in the mechanism taken into account
to predict breakup (e.g. viscous shear, turbulent fluctuations, wall
impact, drag or rotary stress \citep{BK19}).  On the scale of the
aggregate, detailed predictions can be obtained by first principle
structural mechanics
\citep{zaccone2009breakup,conchuir2013mechanism,jiang2020particle} or
by Stokesian dynamics \citep{brady1988stokesian}, with the latter
that, when coupled with models for the inter-particle interactions, is
able to fully characterize the breakup occurrence in terms of critical
stress and fragment size distribution. By such an approach, for
instance, \cite{harada2006dependence} studied the effect of the
internal connectivity on the aggregate breakup. Similarly,
\cite{harshe2012breakage} computed breakup rates and fragment size
distribution in linear flows, and \cite{frungieri2021aggregation}
studied the particle size distribution and the morphological evolution
of a population of colloidal particles at varying shear stress
intensity and physico-chemical properties \citep{Frungieri_2017,
  frungieri2020shear}.

On the macro-scale (i.e., the scale of the equipment), where flow
field heterogeneities and boundary layer phenomena affect the
evolution of the dispersed phase, so called Eulerian-Eulerian
approaches, especially when coupled with population balance models
\citep{marchisio2006role}, are of particular interest, as they can be
conveniently used to promptly compute the breakup dynamics and the
particle size distribution. By such an approach a number of systems of
practical relevance have been investigated, such as emulsions
\citep{lebaz2021modeling}, bubbly flows \citep{syed2018cfd,
  zhang2021cfd, maluta2021effect, lehnigk2022open} and particle
synthesis processes \citep{schikarski2022quantitative}.  However,
despite the wide range of applications, such approaches still rely on
empirical correlations, generally assuming a single phenomenon
(e.g. viscous shear, turbulent fluctuations, surface instability) to
be responsible for breakup, with the overall robustness of the
approach still often in need to be checked against dedicated
experimental campaigns.

More recently, approaches aimed at linking small scale and large scale
phenomena have emerged, most of which adopted an Eulerian-Lagrangian
simulation strategy.  By such an approach, it is possible to study
complex flow fields and treat in a more detailed way the dynamics of
the dispersed phase, especially when the back-reaction of the
particles on the flow is relevant.  In this framework,
\cite{chen2020collision}, considering an homogeneous isotropic
turbulent flow, computed breakup rates in the early stage of an
agglomeration process between adhesive particles, and single events
such as restructuring and breakup by turbulent stresses were also
studied \citep{ruan2020structural, yao2021deagglomeration}.  However,
due to the high computational burden, this approach is limited to
short simulated physical time, low level of turbulence (small
turbulence Reynolds number $Re_\lambda$) and aggregates made by a
small number of primary particles compared to what is typically
observed in experiments \citep{saha2016breakup}.

Turbulence affects particle motion in a distinctive manner, in
particular in the case of inertial particles \citep{BC22}. Inertia
arises when particles have a finite size, and/or a density mismatch
with the suspending medium. Because of inertia, particles show complex
behaviours that notably affect their spatial distribution
\citep{WM93,BecPRL07} and their relative velocity and acceleration
statistics \citep{FP07,BecJP11,scatamacchia2012extreme}.  In
particular, strong inhomogeneities in the particles spatial
distribution emerge, an effect that is maximal when the Stokes number
is of order unity, and becomes negligible in the limits of both small
and large inertia. Moreover, in the case of large inertia, heavy
particles move almost independently of the fluid, hence they may
collide with a large relative velocity. These events -- dubbed {\it
  caustics} -- can cause an substantial increase in the collision rate
\citep{PW16}.

An attempt to understand how the properties of turbulence affect
breakup was undertaken by \cite{GF17} by evolving a population of
particles of variable size in a synthetic turbulent flow.  However,
such a simulation method does not account for turbulence intermittency
which is responsible for the generation of intense hydrodynamic
stresses able to break also the strongest aggregates
\citep{babler2015numerical}.  In this context, \citet{PRE12} computed
the breakup rate of tracer-like aggregates at varying internal
strength in a homogeneous isotropic turbulent flow, assuming particles
to break under the action of the turbulent viscous dissipation
only. Similarly, \citet{debona2014}, by combining a DNS of the
turbulent flow with a Discrete Element Method based on Stokesian
dynamics, estimated the rate of breakup of aggregates addressing at
the same time size and distribution of the formed fragments.

In this work, we study the fragmentation of inertial heavy and light
aggregates in a turbulent flow by combining a direct numerical
simulation of the turbulence with a Lagrangian tracking of the
particles, perfomed within the point-particle approximation of the
Maxey-Riley-Gatignol equations of motion. Likewise to our previous
work \citep{PRE12}, we neglect the internal structure of the
aggregates, the hydrodynamic interactions between them
\citep{zahnow2011particle} and the accumulation of stresses on their
structure \citep{marchioli2015turbulent}, and we assume breakup to
occur whenever the local instantaneous hydrodynamical stresses acting
on the aggregate exceeds a critical value \citep{PRE12,BK19}. We focus
in particular on the role of the Stokes number and of the particle
buoyancy on both shear and drag stresses and we evaluate the rate of
breakup occurrence at varying aggregate strength.

The paper is organised as follows: in Section~\ref{sec:methodology},
we introduce the equation of motion for the inertial particles and the
turbulent flow and we provide some details about their numerical
integration; in Section~\ref{subsec:breakuprate}, we present different
approaches to measure the breakup rate. Results are discussed in
Section~\ref{sec:results}, which are followed by the concluding
remarks in Section~\ref{sec:closeup}.

\section{Methodology}\label{sec:methodology}
We consider a dilute suspension of aggregates described as point-like
spherical particles, which have no feedback on the flow in which they
are suspended, and which have no hydrodynamical interactions between
them. Aggregates have sizes smaller or comparable to the Kolmogorov
scale of the flow $\eta=(\nu^3/\langle \varepsilon \rangle)^{1/4}$,
where $\nu$ is the kinematic viscosity of the fluid and $\langle
\varepsilon \rangle$ is the mean rate of energy dissipation. Both the
particle Reynolds number, defined as $Re_p = 2R\,v_p/\nu$, where $v_p$
is a typical particle velocity and $R$ the particle radius, and the
particle Reynolds number based on the relative particle-fluid
velocity, $2R|v_p - u_f|/\nu$, are small.

\begin{figure}
  \centerline{\includegraphics[width=0.6\linewidth]{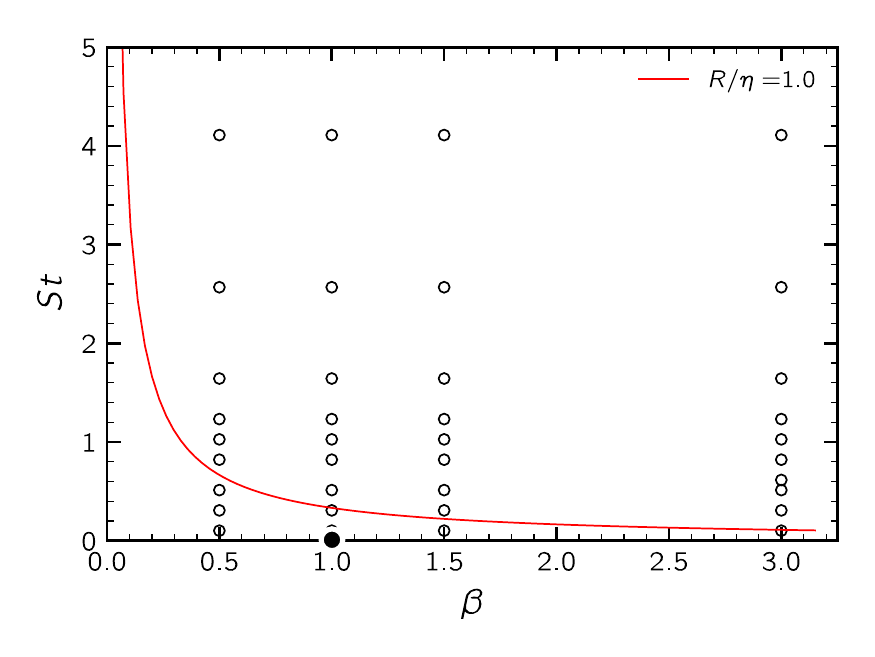}}
  \caption{Parameters space of the particles used
    in our numerical experiments. 
    The red line represents the locus of constant $R/\eta$. We analyzed $128k$ trajectories per each single $(\beta,St)$ choice.
    Tracers are represented by the black dot. Trajectories were evolved using a second order algorithm, with a time step $dt$ such that $\tau_{\eta}
    \simeq 100dt$.}
\label{fig:dataset}
\end{figure}

Aggregate trajectories are obtained by evolving a minimal formulation
of the original Maxey-Riley-Gatignol equations of motion
\citep{MR1983,Gat1983} in which the pressure force, the added mass and
the Stokes drag are kept into account, and which has been frequently
used to describe the motion of small, rigid, spherical particles in
unsteady flows \citep{JFM2010pair}. Such equations read as:
\begin{eqnarray}
  \frac{d \mathbf{X}}{dt} &=&
  \mathbf{V}(\mathbf{X},t)\,, \label{eqmotion} \\ \frac{d
    \mathbf{V}}{dt}&=&\beta\frac{D
    \mathbf{u}(\mathbf{X},t)}{Dt}+\frac{\mathbf{u}(\mathbf{X},t) -
    \mathbf{V}(\mathbf{X},t)}{\tau_p}\,, \label{eqmotion2}
\end{eqnarray}
where $\mathbf{V}$ is the particle velocity, $\mathbf{X}$ is the
particle position and $\mathbf{u}$ is the undisturbed fluid velocity
at the particle position.  It is apparent that only two dimensionless
parameters govern the particle motion: the density ratio $\beta$
defined as $\beta=\frac{3\rho_f}{\rho_f+2\rho_p}$ where $\rho_p$ and
$\rho_f$ represent the particle and fluid density, respectively, and
the Stokes number $St=\tau_p/\tau_\eta$, where the particle relaxation
time is defined as $\tau_p = R^2/(3\beta \nu)$, being $R$ the particle
radius. The Kolmogorov time scale of the flow, entering the definition
of the Stokes number, is $\tau_{\eta}=(\nu/\langle \varepsilon
\rangle)^{1/2}$.  From the definition of $\beta$, it is clear that
neutrally buoyant particles have $\beta=1$, extremely light particles
have $\beta \to 3$, while heavy particles have $\beta \to 0$.  We
track the motion of particles at varying inertia, by changing both
their buoyancy parameter and their Stokes number as schematically
illustrated in Fig.~\ref{fig:dataset}.

The fluid velocity $\mathbf{u}$ is evolved according to the
incompressible Navier-Stokes (NS) equations reading as:
\begin{equation}\label{NS}
  \frac{\partial \mathbf{u}}{\partial t} + \mathbf{u}\cdot \nabla \mathbf{u}=-\frac{\nabla p}{\rho_f}
  + \nu \nabla^2 \mathbf{u} + \mathbf{F}, \qquad \nabla \cdot \mathbf{u}=0\,.
\end{equation}
A steady, statistically homogeneous and isotropic turbulent flow is
obtained by adding to the NS equations a forcing term $\mathbf{F}$
injecting energy in the first low-wave number shells and keeping
constant their spectral content \citep{JFM2010pair}. The kinematic
viscosity is chosen in such a way that the Kolmogorov length scale
equals the grid spacing $\eta \simeq \delta x$. By doing so, a good
resolution of the small-scale velocity fluctuations is obtained. At
the steady state, the energy input balances the mean kinetic energy
dissipation such that $\langle \mathbf{F} \cdot \mathbf{u} \rangle
\simeq \langle \varepsilon \rangle$. The Navier Stokes equations are
solved on a $512^3$ cubic grid with periodic boundary conditions and a
Taylor–Reynolds number $Re_\lambda \simeq 185$. In
Table~\ref{table:flow} the main characteristics of the flow are
reported. Further numerical details on both the Eulerian and
Lagrangian approaches can be found in the work by \cite{JFM2010pair}.
\begin{table}
    \caption{Parameters of the DNS simulation. Microscale Reynolds number
  $Re_\lambda$, root-mean-square velocity $u_{\rm rms}$, mean energy
  dissipation $\varepsilon$, kinematic viscosity $\nu$, Kolmogorov
  scale $\eta = (\nu^3/\langle \varepsilon\rangle)^{1/4}$, integral scale $L$,
  Eulerian large-eddy turnover time $T_E = L/u_{\rm rms}$,
  Kolmogorov timescale $\tau_\eta=(\nu/\langle \epsilon \rangle)^{1/2}$, grid
  spacing $\Delta x$, number of grid points $N$, simulation time $t_{\rm s}$. \label{table:flow}}
\begin{center}
    \begin{tabular}{ccccccccccc}\hline
     $Re_\lambda$ & $u_{rms}$ & $\varepsilon$ & $\nu$ & $\eta$ & $L$ &
      $T_E$ & $\tau_\eta$ & $\Delta x$ & $N^3$ & $t_{\rm s}$\\ 185 & 1.4 & 0.94 &
      0.00205 & 0.010 & $\pi$ & 2.2 & 0.047 & 0.012 & $512^3$ & 13.2\\ \hline
    \end{tabular}
    \end{center}

    \end{table}

\subsection{Hydrodynamic stresses}
\label{sec:stresses}
Once a statistically steady state condition in the flow is reached,
particles are released at random in the fluid, and the time-series of
the hydrodynamic stresses acting on them is tracked. We assume
aggregates to be brittle objects that instantaneously respond to the
external stress and that undergo breakup as soon as their critical
resistance is exceeded by the total fluid dynamic stress acting on
them.

Following \citet{Kusters1991} and \citet{BK19}), two hydrodynamic
stresses are deemed as responsible for the breakup of the aggregate,
namely the shear stress $\sigma_\varepsilon$, due to the fluid
velocity gradients at the particle position, and the drag stress
$\sigma_{St}$, due to the slip velocity between the particle and the
underlying flow. The first is evaluated along the particle trajectory
$\mathbf{X}$ as:
\begin{equation}
\label{eq:shearstress}  
    \sigma_{\varepsilon}(\mathbf {X},t)= \mu \sqrt{\frac{2}{15}\frac{\varepsilon(\mathbf {X})}{\nu}}\,,
\end{equation}
where $\mu= \nu \rho_f$ is the dynamic viscosity of the flow, and where the local turbulent dissipation rate was computed as $\varepsilon=2\nu e_{ij}e_{ij}$, being $e_{ij}=1/2 (\nabla_j
u_i + \nabla_i u_j)$ the symmetric part of the velocity gradient tensor
$\nabla_j u_i$.

The drag stress is instead evaluated as:
\begin{equation}
\label{eq:dragstress}  
    \sigma_{St} (\mathbf {X},t)=\mu\frac{3\vert{\bf u}(\mathbf{X})-{\bf V}(\mathbf{X})\vert}{2R}\,.
\end{equation}
where we have assumed, as above, that the stress is isotropic and that
the aggregate has a spherical shape (see also \cite{BK19} for a
discussion). Therefore the drag stress acting on an aggregate can be
simply computed as the ratio between the drag force $6\pi\mu R
\vert{\bf u}-{\bf v}\vert$ and the surface area $4 \pi R^2$.
Following \cite{Kusters1991}, the two stresses are added up linearly
$\sigma_{tot}=\sigma_{St}+\sigma_\epsilon$, thus assuming that the two
stresses propagate instantaneously across the aggregate bond network
and that their load on the structure can be superimposed.

\subsection{Breakup rate measurements}
\label{subsec:breakuprate}

\begin{figure}
\centering
\includegraphics[width=0.6\linewidth]{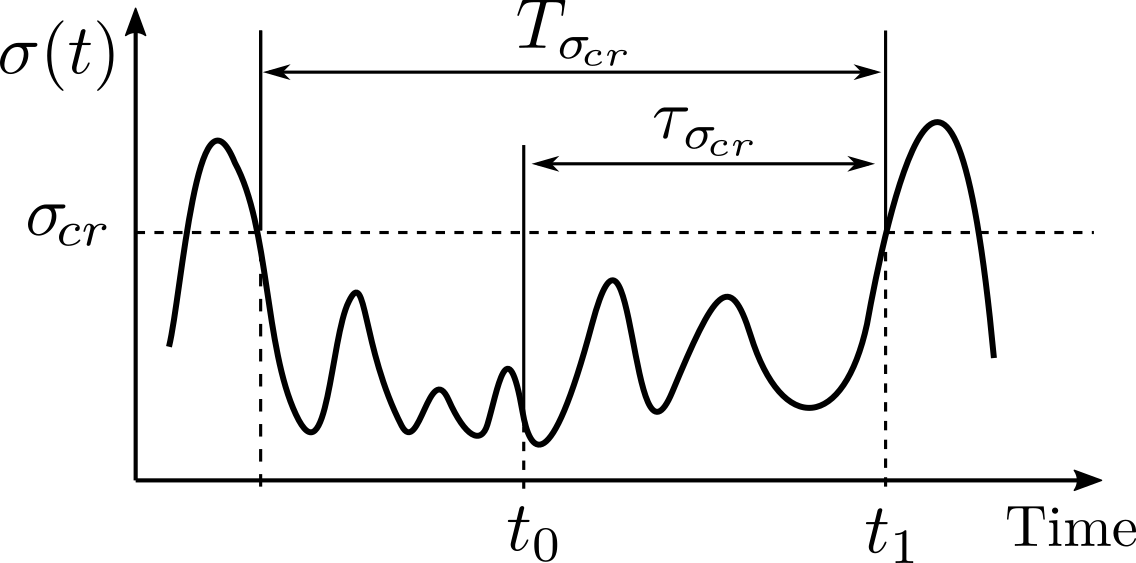}
\caption{\label{fig:exit_diving} Schematic of the total hydrodynamic stress along a particle trajectory. The particle is released at time $t_0$ and breaks up at time $t_1$. The time-lags $\tau_{\sigma_{cr}}$ and $T_{\sigma_{cr}}$ are the exit time and the diving time, respectively. }
\end{figure}
Our work is aimed at evaluating particle breakup rates at varying
inertia. In our model, an individual aggregate undergoes breakup
instantaneously when the total hydrodynamic stress acting on it
exceeds a critical threshold value $\sigma_{cr}$, representing the
aggregate mechanical strength.

Figure~\ref{fig:exit_diving} illustrates the simple modeling framework
we use. The aggregate is released at a random time $t_0$ and samples
the flow until it experiences for the first time a hydrodynamic stress
exceeding its resistance $\sigma_{cr}$ (this, in the schematics of
Fig.~\ref{fig:exit_diving}, occurs at $t_1$). The time-lag between
release at $t_0$ and breakup at $t_1$ identifies the exit time
$\tau_{\sigma_{cr}}$. The breakup rate is therefore calculated as the
inverse of the mean exit time, i.e., as:
\begin{align}\label{eq:f_exact}
f_{\sigma_{cr}} \equiv \frac{1}{\langle \tau_{\sigma_{cr}} \rangle}\,.
\end{align}
Notice that within this modelling framework, situations where the
hydrodynamic stress exceeds the critical stress already at the point
of release are ignored when computing the average in
Eq.~\eqref{eq:f_exact}. The reason for this is that breakup events
that occur right at the point of release would be governed by the
frequency of aggregate release and not by the timescale of the
turbulent fluctuations.

The picture presented in Fig.~\ref{fig:exit_diving} can be used to
identify a second time scale. For this, we notice that the
hydrodynamic stress experienced by the aggregate during its lifetime
is part of the Lagrangian time series of a particle with the same
kinematic characteristics (Stokes number and density parameter) as the
aggregate. Since this Lagrangian time series exists throughout the
lifetime of the flow, the hydrodynamic stress can also be traced
backwards in time, as depicted in Fig.~\ref{fig:exit_diving}. Hence,
by considering the Lagrangian time series of which the aggregate
released at $t_0$ sampled the segment between $t_0$ and $t_1$, we can
identify a new time scale, called diving time $T_{\sigma_{cr}}$, as
the time-lag between the last down-crossing of $\sigma_{cr}$ before
$t_0$ and the first up-crossing of $\sigma_{cr}$ after $t_0$.  This
time scale $T_{\sigma_{cr}}$ allows us for deriving a few additional
relationships to express or approximate the breakup rate. At first, it
can be shown that for a statistically stationary flow the breakup rate
defined in Eq.~\eqref{eq:f_exact} can also be expressed through the
variance of the diving time as \citep{PRE12}:
\begin{align}\label{eq:T1_T2}
f_{\sigma_{cr}}=\frac{2\langle T_{\sigma_{cr}}\rangle}{\langle T_{\sigma_{cr}}^2\rangle}
\end{align}
Moreover, we can define a proxy-breakup rate by using the diving time
instead of the exit time. The proxy-breakup rate simply reads as:
\begin{align}\label{eq:f_proxy}
\tilde f_{\sigma_{cr}}=\frac{1}{\langle T_{\sigma_{cr}}\rangle}	
\end{align}
For a stationary stochastic process, the above can be calculated by
the frequency of upward-crossing of the the threshold $\sigma_{cr}$
divided by the fraction of time the trajectory stays below the
threshold. The former can be expressed by means of the Rice theorem,
whereas the latter follows from the probability density function (PDF)
of $\sigma_{tot}(t)$ \citep{lindgren2019gaussian}. By doing so, the
proxy-breakup rate can be computed as:
\begin{align}\label{eq:f_loginov}
\tilde f_{\sigma_{cr}}=\frac{
\int_0^\infty d\dot\sigma_{tot}\, \dot\sigma p_2(\sigma_{cr},\dot\sigma_{tot})
}{
\int_0^{\sigma_{cr}} d\sigma_{tot} \, p(\sigma_{tot})
}
\end{align}
where $p_2(\sigma_{tot},\dot\sigma_{tot})$ is the joint PDF of the
total stress and of its time derivative, and where $p(\sigma_{tot})$
is the PDF of $\sigma_{tot}$. Approximating the breakup rate using the
Rice formula was first proposed by \citet{Lo85}. Accordingly,
Eq.~\eqref{eq:f_loginov} is referred to as Loginov's approximation.

For large values of the threshold $\sigma_{cr}$, the proxy-breakup
rate based on the diving time (Eq.~\eqref{eq:f_proxy}) approaches the
exact breakup rate based on the exit time (Eq.~\eqref{eq:f_exact}):
\begin{align}\label{eq:largethreshold}
\langle\tau_{\sigma_{cr}}\rangle \sim   
\langle T_{\sigma_{cr}}\rangle \quad\textrm{for}\quad 
\sigma_{cr}\gg\langle\sigma_{tot}\rangle
\end{align}
To understand this, let us consider a long timeseries of length $t_L$, with $t_L$ being much larger than the large eddy turnover time $T_E$.   
Along this time series, the threshold $\sigma_{cr}$ is
crossed $N$-times in the upward direction. Since the threshold is very large, the total time the timeseries spends above $\sigma_{cr}$ is
negligible and the mean diving time follows as $\langle T\rangle =
t_L/N$. To connect this to the exit time, we assume that the
up-crossings of $\sigma_{cr}$ are distributed randomly along the
trajectory. Considering that the timespan between two consecutive upcrossings is large compared to the large eddy turnover time $T_E$, this assumption is reasonable.

Next to the $N$ upward-crossing events along the trajectory, there is
now also another event i.e., the release of the aggregate at a random
time $t_0$, such that in total there are $N+1$ events along the
trajectory. Thus, the mean exit time follows as
$\langle\tau\rangle=t_L/(N+1)$.  For a stationary flow, the length
$t_L$ and the number of upward-crossings $N$ are large such that
$t_L/N \approx t_L/(N+1)$, finally rationalizing the result of
Eq.~\eqref{eq:largethreshold}.

\section {Results and discussion}
\label{sec:results}

\subsection{Hydrodynamic stress statistics}
\begin{figure}
  \centerline{\includegraphics[width=0.7\linewidth]{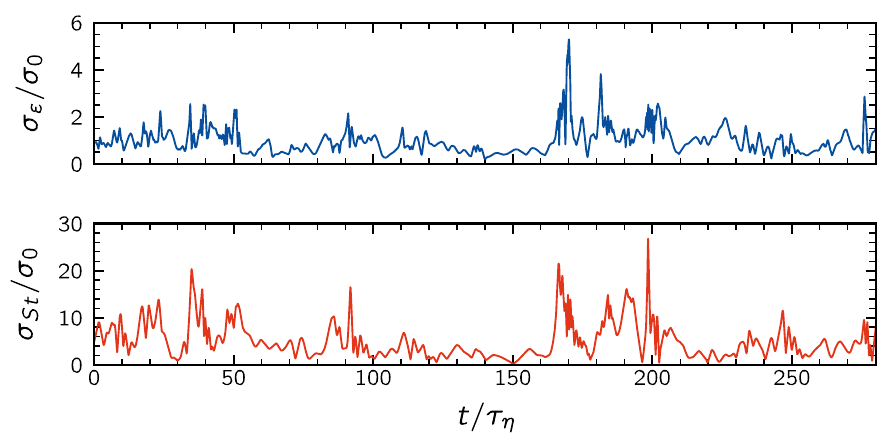}}
  \caption{Time series of the shear stress $\sigma_\varepsilon$ (top
    panel) and drag stress $\sigma_{St}$ (bottom panel) for a sample
    aggregate with Stokes number $St=1.64$ and buoyancy parameter
    $\beta=3$. The $x$ axis is normalized by the Kolmogorov time scale
    of the flow. The $y$-axis is normalized by the average shear
    stress $\sigma_0$ experienced by tracer particles ($\beta=1$,
    $St=0$).}
\label{fig:signals}
\end{figure}

To study the statistics of the hydrodynamic stresses experienced by
inertial aggregates, we start by consider them as infinitely strong
and moving in the flow field without undergoing
breakup. Figure~\ref{fig:signals} shows the time series of the shear
stress $\sigma_\varepsilon$ and of the drag stress $\sigma_{St}$,
recorded along the trajectory of a heavy aggregate with inertia values
$\beta=0.5$ and $St=1.64$. We observe that both stresses exhibit
strong fluctuations, each with its own dynamics. Indeed, the first is
determined by the way the particle sample the different regions of the
flow, whereas the second comes from the fluid-particle slip
velocity. For the chosen set of inertia parameters, we also observe
that the drag stress dominates over the shear stress along the whole
particle trajectory.
\begin{figure}
  \centerline{\includegraphics[width=0.6\linewidth]{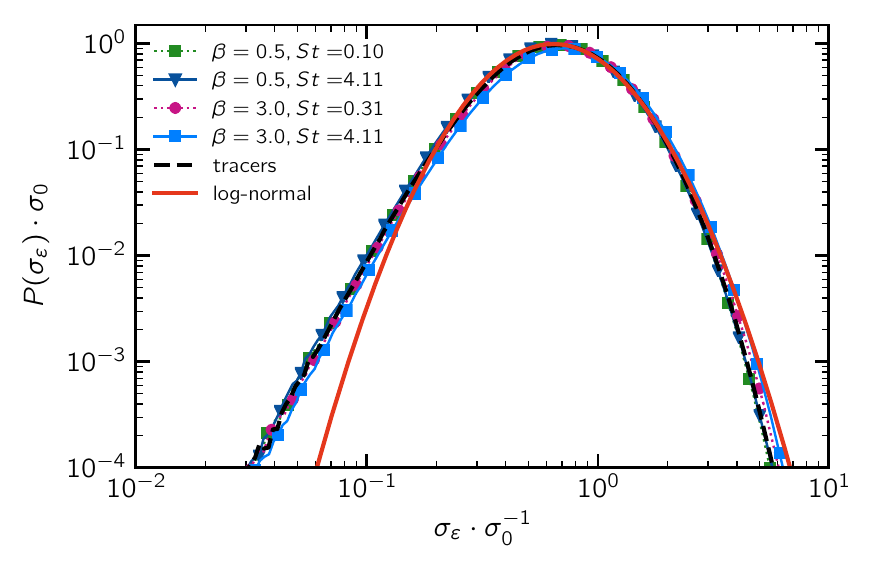}}
  \caption{Probability density function of the shear stress at varying
    particle inertia. The red dotted line represents a log-normal
    distribution $P(\sigma_0,0.55\sigma_0)$, with $\sigma_0$ being the
    average $\sigma_\varepsilon$ experienced by tracer particles.}
\label{fig:PDFeps}
\end{figure}
Figure~\ref{fig:PDFeps} reports the probability density function of
the shear stress for two buoyancy parameters ($\beta=0.5$ and
$\beta=3.0$, corresponding to heavy and light particles, respectively)
and for the smallest and largest Stokes values here investigated. The
plot also reports the distribution of the shear stress experienced by
tracer particles, for which $\beta=1$ and $St \to 0$.  It is apparent
that, for all combination of $\beta$ and $St$, the curves are very
similar and, more remarkably, regardless of their inertia, all
particles sample the shear stress field exactly as tracer particles
do, as it is made apparent by the fairly good overlap of the
distributions within the whole stress range.  Furthermore, the shear
stress $\sigma_\varepsilon$ appears to be well fitted, at least for
the central part of the value range, by a log-normal distribution
peaking at the average value of the shear stress $\sigma_0$
experienced by tracers.

\begin{figure}
\centering
\includegraphics[width=0.95\linewidth]{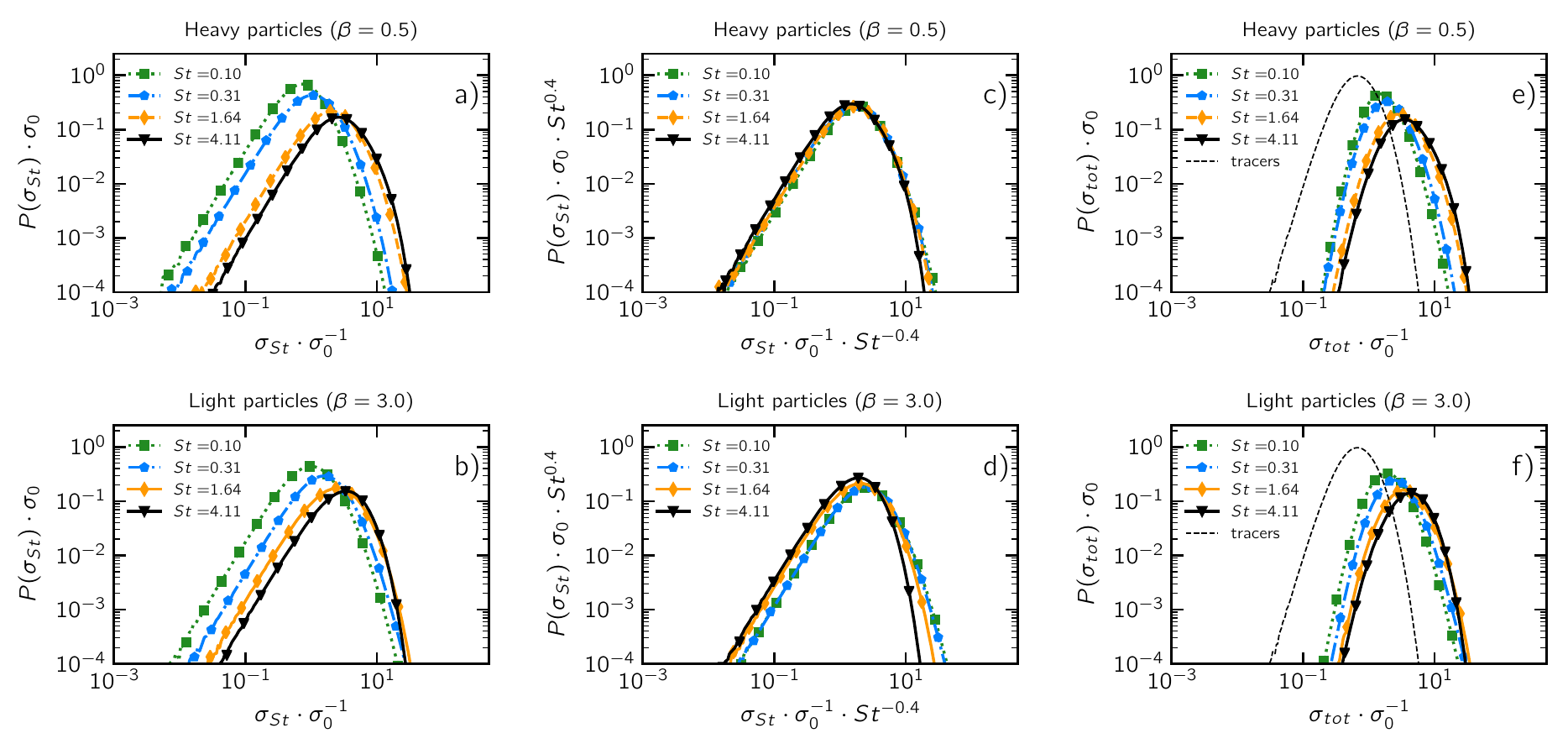}
\caption{Probability density function of the fluid dynamic stresses
  for heavy and light particles at varying Stokes number. Panel a) and
  b) report the PDFs of the drag stress. In panel c) and d) the same
  PDFs are rescaled by $St^{0.4}$. Panel e) and f) report the total
  stress distributions. In these two panels, the black dotted line is
  the PDF of the shear stress experienced by tracers.}
\label{fig:PDF}
\end{figure}

\begin{figure}
\centering
\includegraphics[width=0.45\textwidth]{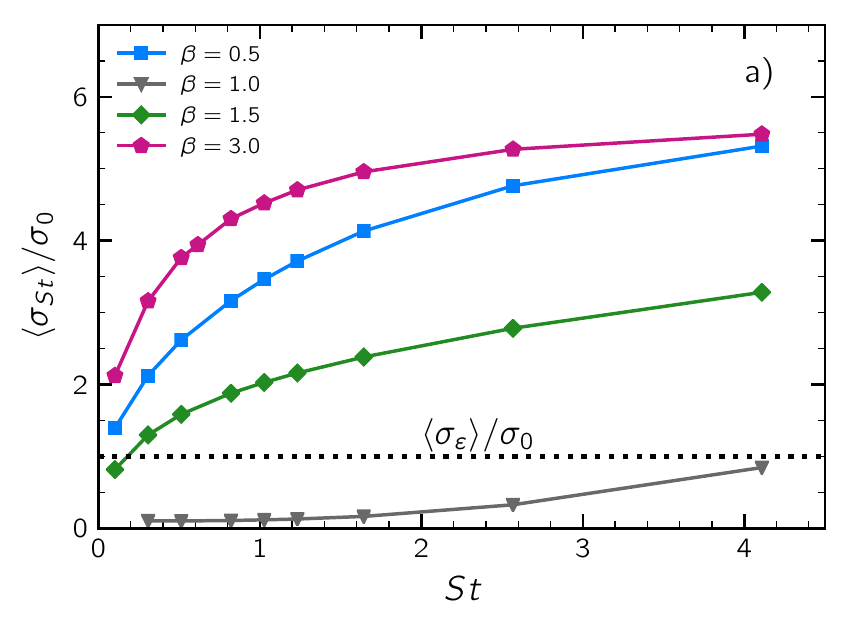}
\includegraphics[width=0.49\textwidth]{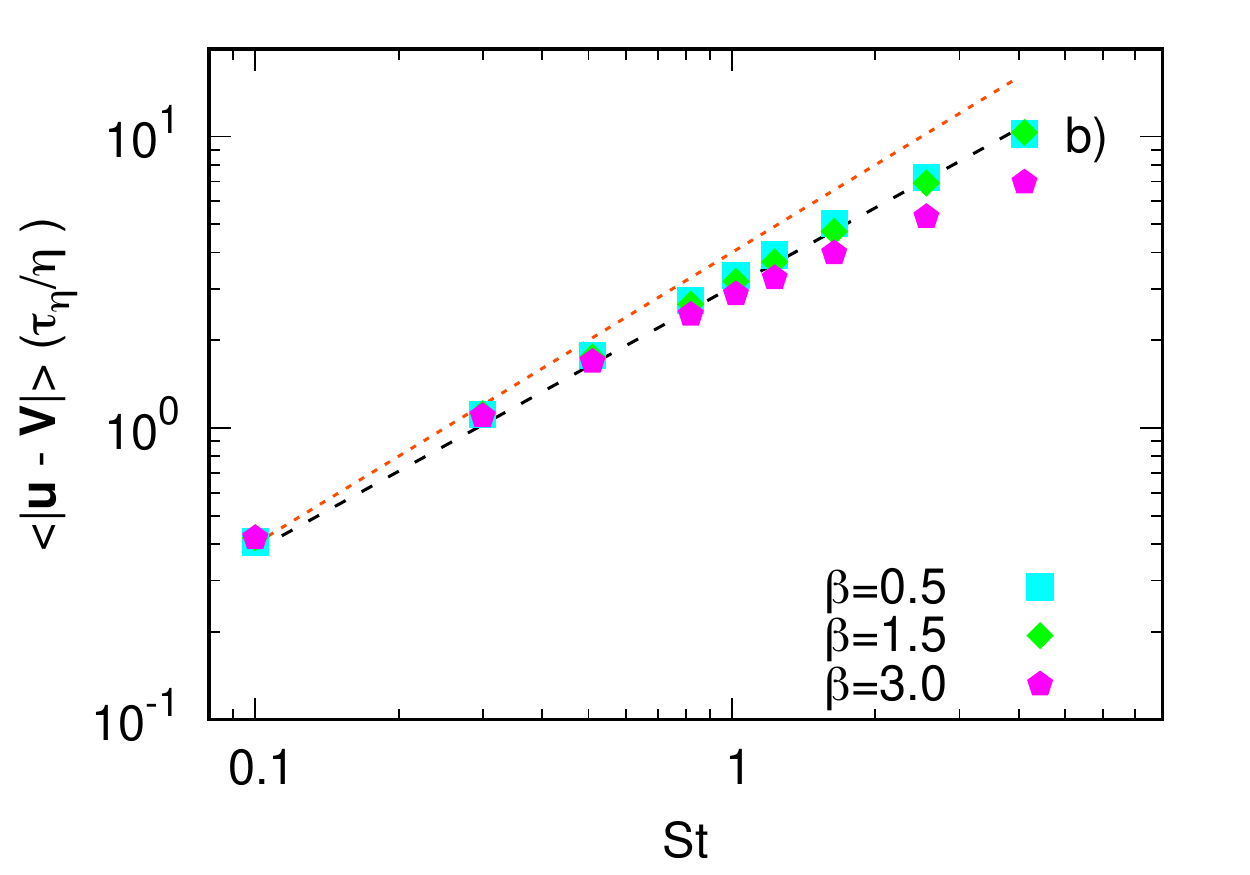}
\caption{a) Drag stress for various $\beta$ values as a function of
  $St$. The dotted line reports the average $\sigma_\varepsilon$
  experienced by tracers. b) Log-log plot of the fluid-particle slip
  velocity as a function of the Stokes number, for various $\beta$
  values. The slip velocity is made dimensionless by the Kolmogorov
  time $\eta$ and length scale $\tau_\eta$; for $\beta=0.5$ and
  $\beta=1.5$, the slip velocity have been shifted vertically for
  plotting purposes. The red dotted line shows the theoretical scaling
  in the limit of small Stokes number $\langle \lvert{\bf u} -{\bf
    V}\rvert\rangle \propto St$; while the black dashed line shows the
  scaling $\langle \lvert{\bf u} -{\bf V}\rvert \rangle \propto
  St^{0.9}$.}
\label{fig:dragstress}
\end{figure}

Figures~\ref{fig:PDF}a) and b) show the PDF of the drag stress for
heavy and light particles, respectively, at varying Stokes number. In
these two panels the PDFs are shown in dimensionless coordinates,
without scaling.  The average values of the drag stress for a larger
set of particle properties are shown in Fig.~\ref{fig:dragstress}a),
where it is apparent that the average drag stress increases with the
Stokes number. To better understand the general behavior, we recall
that the drag stress is proportional to the slip velocity $\lvert {\bf
  u}-{\bf V}\rvert$ (see Eq.~\eqref{eq:dragstress}).

In Fig.~\ref{fig:dragstress}b), we show the normalised average slip
velocity as a function of both the Stokes number and the buoyancy
parameter: at small Stokes number, we observe that the mean slip
velocity scales linearly for all the investigated buoyancy parameters,
in agreement with the Maxey approximation for weak inertia
\citep{Maxey1987}. Indeed, starting from Eq.~\eqref{eqmotion2}, in the
limit of $St \rightarrow 0$, the slip velocity becomes:
\begin{equation}
{\bf u} -{\bf V} \equiv \tau_p \left(d{\bf V}/dt - \beta D{\bf u}/Dt \right) \simeq \tau_p (1-\beta) a_f\,,
\end{equation}
where we have assumed that in the limit of small Stokes numbers the
aggregate acceleration and the fluid acceleration along the aggregate
trajectory are equal, i.e., $d{\bf V}({\bf X})/dt \simeq D{\bf u}({\bf
  X})/Dt \simeq a_f({\bf X})$. Hence, in the weak inertia limit,
$\langle \lvert{\bf u} -{\bf V}\rvert \rangle \propto St$
\citep{Boffetta2007} as shown by the dotted line in
Fig.~\ref{fig:dragstress}b.

However, we also observe that the mean slip velocity can be described
fairly well by the power-law scaling $\langle \lvert{\bf u} -{\bf
  V}\rvert \rangle \simeq St^{\alpha}$, with $\alpha = 0.9$: in other
words, an {\it effective} and {\it empirical} exponent, $\alpha$, can
be introduced that models the slip velocity behaviour over a wide
range of Stokes numbers (up to $St \sim O(1)$). Such an exponent is
found to be independent, or weakly dependent, on the buoyancy
parameter.  Recalling that the drag stress $\sigma_{St} \propto
\lvert{\bf u} -{\bf V}\rvert/R$, and that $R \propto O(St^{1/2})$, we
finally obtain that $\sigma_{St} \propto St^{0.4}$.  If we apply this
scaling to the probability distribution functions of the drag stress
of Figs.~\ref{fig:PDF}a,b), we observe that the distributions collapse
onto a single curve for both light particles (up to $St \approx 1.64$)
and heavy particles (up to $St \approx 4.11$) over a quite wide range
of $\sigma_{St}$ (see Figs.~\ref{fig:PDF}c,d).  We also note that the
left tail of the drag stress is well described by a $\chi^2$
distribution with 3 degrees of freedom. We consider this an outcome of
the Gaussian dynamics of the smallest stress contributions. To
summarise, the collapse of the left tails of the PDFs is clearly a
consequence of the fact that small fluctuations of the slip velocity
are proportional to the first moment of the distribution, as expected;
moreover, these small fluctuations occur in the smooth regions of the
flow.

The panels e) and f) in Fig.~\ref{fig:PDF} show the probability
distribution functions of the total stress for heavy and light
particles. The distribution of the shear stress experienced by
tracers, which, as already discussed, is very similar to the one of
the other particle families, is also reported for comparison.  The
shifting of the curves towards the right at increasing $St$ makes it
apparent that, as inertia is increased, particles experience larger
drag stresses. However, at Stokes number $\approx 4$, a drop down in
the probability of the largest stresses can be observed, especially
for light particles.

\begin{figure}
\centering
\includegraphics[width=0.45\linewidth]{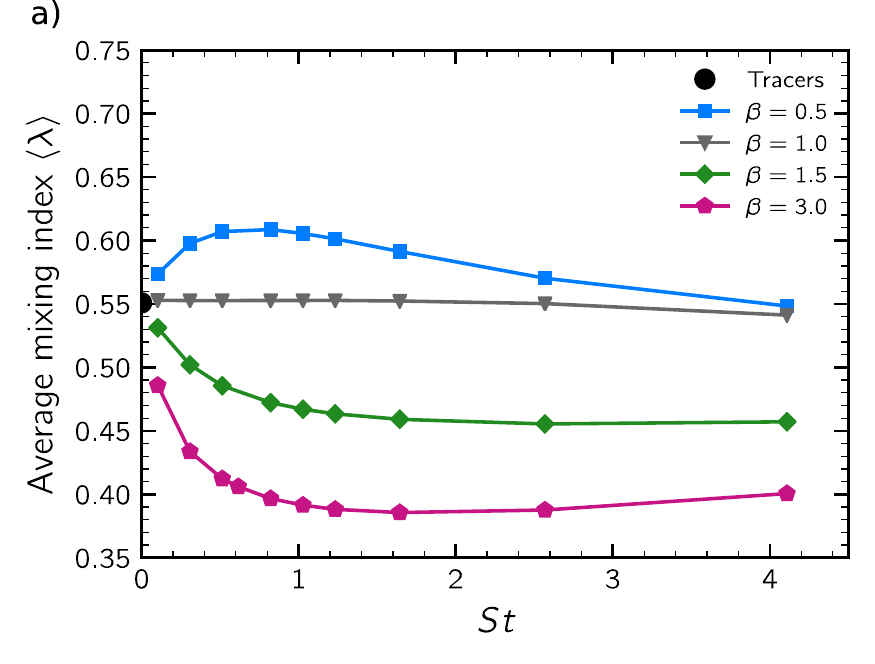}
\includegraphics[width=0.51\linewidth]{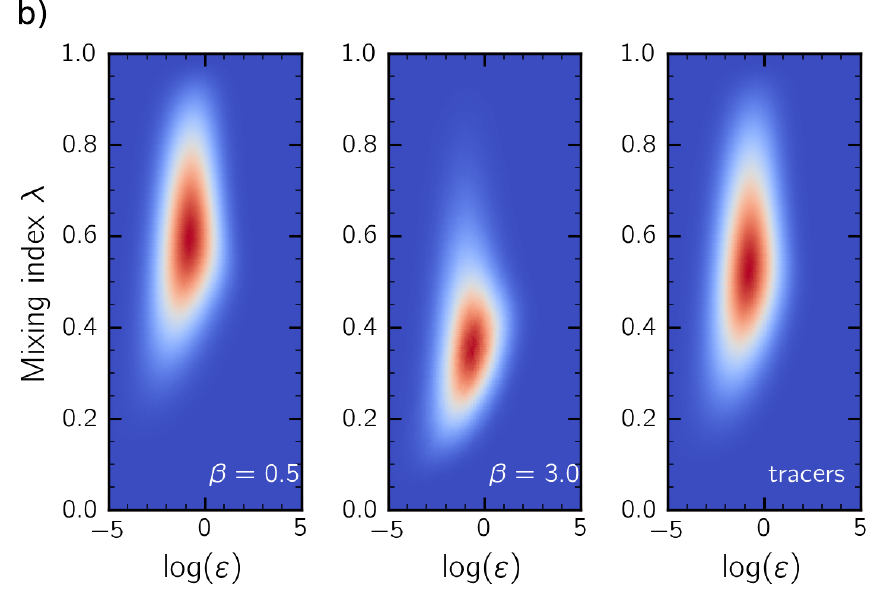}
\caption{a) Average mixing index seen by aggregates as a function of both $St$ and $\beta$. The black circle indicates the tracer behavior. b) Bivariate distribution function of the turbulent dissipation rate and mixing index for heavy aggregates ($\beta=0.5$), light aggregates ($\beta=3$) and tracers. Stokes number is approximately equal to 1 for both the light and the heavy aggregate class.}
\label{fig:mixind}
\end{figure}

Even if particles experience the shear stress space in a way similar
to the one of tracers, when looking at the statistics of the
experienced flow topology, important differences emerge. We
characterize the flow topology along particle trajectories according
to the mixing index $\lambda$, defined as:
\begin{equation}
\lambda=\frac{\sqrt{II_\mathbf{E^\infty}}}{\sqrt{II_\mathbf{E^\infty}}+\sqrt{II_\mathbf{\boldsymbol{\Omega}^\infty}}}
\end{equation}
where $II_{\mathbf{E}^\infty}$ is the second invariant of the
rate-of-strain tensor $\mathbf{E}^\infty$ and where
$II_\mathbf{\boldsymbol{\Omega}^\infty}$ is the second invariant of
the vorticity tensor $\boldsymbol{\Omega}^\infty=0.5\left(\nabla
\mathbf{u}^\infty-\nabla \mathbf{u}^{\infty, T}\right)$. The mixing
index has a 0--1 range, with 0 indicating a pure rotational motion,
and 0.5 and 1 indicating pure shear and pure elongational flow,
respectively.  Figure~\ref{fig:mixind}a) reports the average mixing
index $\lambda$ seen by the aggregates for different Stokes numbers
and buoyancy parameter $\beta$. The black dot at $\lambda=0.55,St=0$
indicates the behavior of tracers. The plot makes apparent that,
regardless of the $\beta$ parameter, small Stokes aggregates behave
very similarly to tracer aggregates with an average mixing index
fairly close to 0.55. However, as the Stokes number increases,
qualitative differences can be observed. Neutrally buoyant aggregates
($\beta=1$) keep behaving as tracer aggregates, with the average
$\lambda$ almost constantly equal to 0.55 at increasing $St$. On the
contrary, heavy and light aggregates show two different behaviors: the
heavy aggregates ($\beta=0.5$) present a maximum in $\langle \lambda
\rangle$ for $St \approx 1$ and converge to the tracers behaviour
again as Stokes increases.

Light aggregates ($\beta=1.5$ and $\beta=3$), on the contrary, have a
minimum $\langle \lambda \rangle$ at $St \approx 1$, which is kept
almost constantly up to $St=4$.  This confirms what already pointed
out by other researchers \citep{calzavarini2008dimensionality,
  calzavarini2008quantifying}, who observed that light aggregates
preferentially sample the vortical regions of the flow (i.e., lower
$\lambda$ regions), whereas heavy aggregates are ejected from vortical
regions and tend to sample the higher strain regions (larger
$\lambda$).  This behavior, referred to as \emph{turbulence induced
  segregation}, is here observed to be particularly important at $St
\approx 1$, in line with what reported by \citet{bec2005clustering}.
Finally, to make this behavior more apparent, Fig.~\ref{fig:mixind}b)
reports the joint probability distribution functions of the mixing
index and turbulent dissipation rate $\varepsilon$ for two different
aggregates classes, with same Stokes number ($St \approx 1$) and
different $\beta$ parameters, and for tracers. It is again apparent
that all aggregates sample almost equally the $\varepsilon$ space, but
quite differently the $\lambda$ space: heavy aggregates experience the
larger values of the $\lambda$ range and behave more similarly to
tracers, except for the lower amplitude of the fluctuations in the
$\lambda$ direction; light aggregates ($\beta=3$) sample the lower
value range of the mixing index and experience smaller
fluctuations. This confirms that are more preferably trapped in the
vortical regions of the flow.

\subsection{Breakup statistics}
\begin{figure}\centering
\includegraphics[scale=0.55]{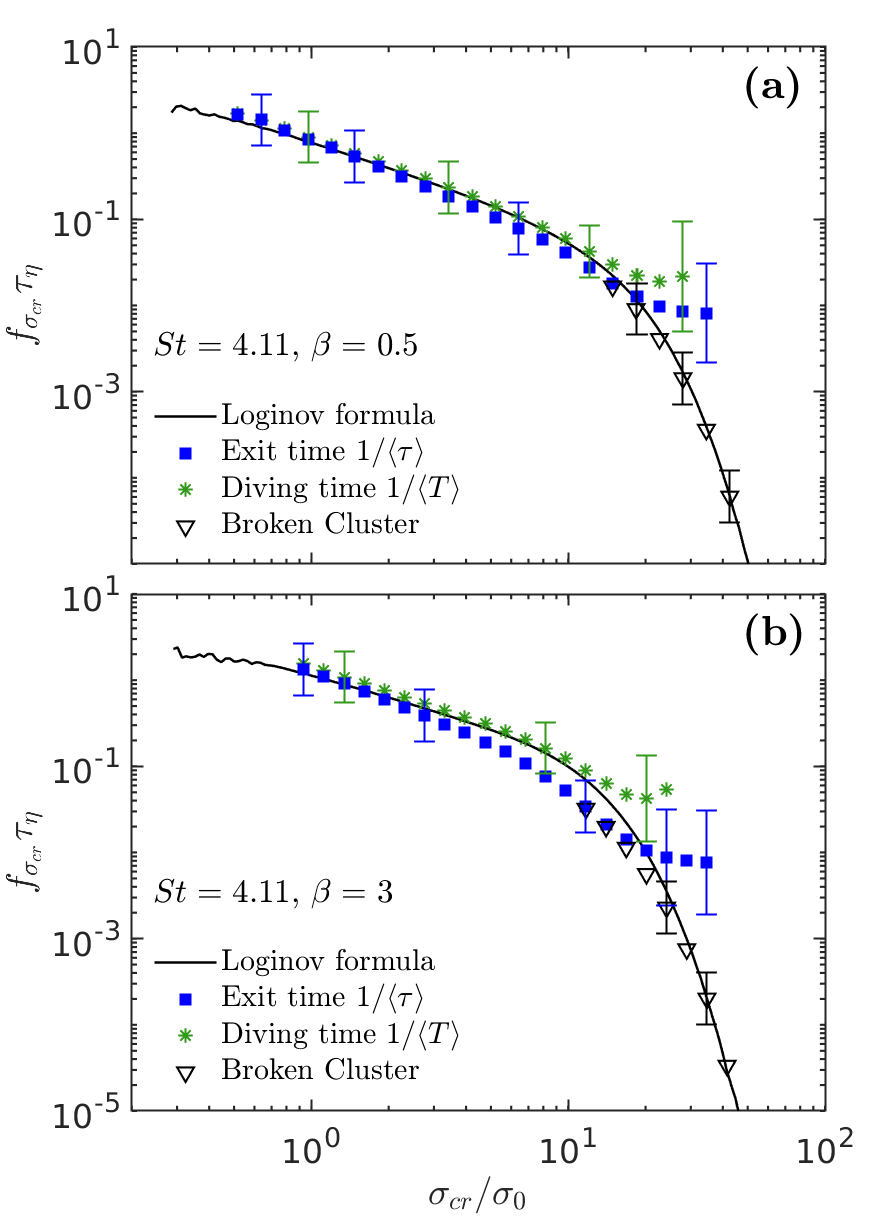}
\caption{\label{fig:breakup22_63} Breakup rate versus the threshold
  stress for aggregates with $St=4.11$ and (a) $\beta=0.5$ and (b)
  $\beta=3$ measured by the exact expression based on the exit time
  (square symbols) and the approximations based on the diving time
  (star symbols), broken clusters (triangles) and the Loginov formula
  (lines). The horizontal axis is normalized by the mean shear stress
  $\sigma_0$, whereas the vertical axis is normalized by the
  Kolmogorov time scale $\tau_\eta$. For the sake of clarity, error
  bars are shown for few data points only.  }
\end{figure}
Figure~\ref{fig:breakup22_63} shows the breakup rate as a function of
the threshold stress for two particle families with same Stokes number
($St=4.11$) and different buoyancy parameters, namely $\beta=0.5$
(panel a) and $\beta=3$ (panel b). The threshold stress on the
horizontal axis is normalized by the mean shear stress $\sigma_0$ and
the breakup rate is normalized by the Kolmogorov time scale
$\tau_\eta$.  Although the data shown refers to particle families on
the outskirts of our dataset (see Fig.~\ref{fig:dataset}), the graph
reflects the general trend of the breakup rate and allows us for
discussing its characteristics and the different approaches used for
measuring it.  As the total stress is calculated from the three
components of the slip velocity and from the spatial derivatives of
the fluid velocity, that are both quantities that fluctuate along a
particle trajectory, the exit time has to be considered as an outcome
of a combination of random variables.  For small threshold values,
this combination leads to a breakup rate following a power-law
behavior: in this regime, in fact, breakup events are controlled by
hydrodynamic stresses that are close to the mean stress and occur in
the smooth regions of the flow, where the stress statistics are
Gaussian distributed.

However, as the threshold increases, the breakup rate has a sudden
super-exponential drop-off. This behaviour is the consequence, on the
contrary, of the rare turbulent events that are vigorous enough to
break strong aggregates, whose occurrence is controlled instead by
turbulent intermittency.
\begin{figure}\centering
\includegraphics[scale=0.55]{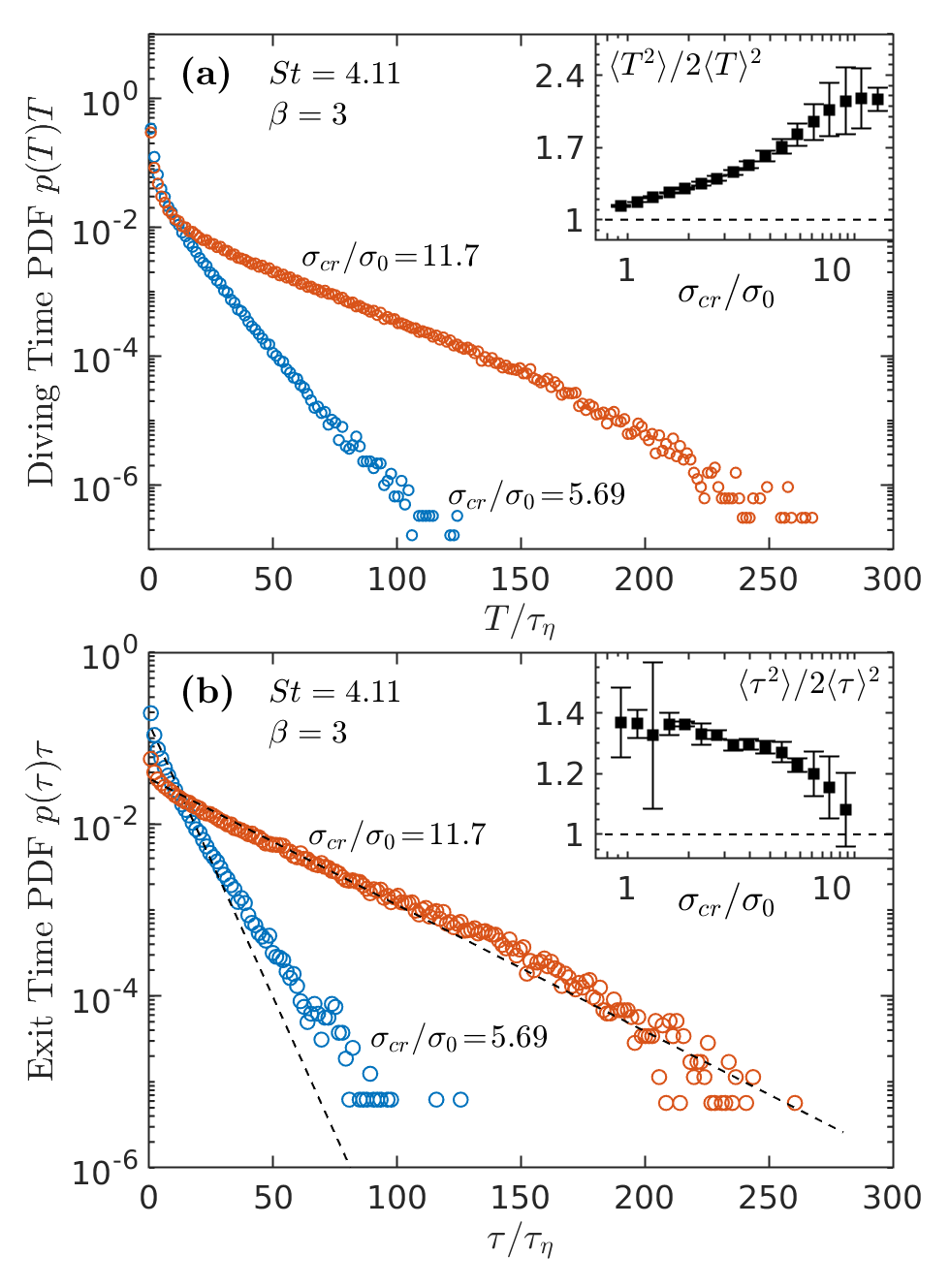}
\caption{\label{fig:exitdiving_fam63} Probability density function for
  aggregates with $St=4.11$ and $\beta=3$ of (a) the diving time and
  (b) the exit time at two critical stress values. The dashed lines in
  (b) show exponential distributions with an expected value equal to
  the mean exit time for the given threshold. The insets show the
  normalized second order moment of the diving time and the exit time
  as a function of the critical stress.}
\end{figure}

Regarding the different approaches for measuring the breakup rate, it
is observed that for small threshold values the exact breakup rate
based on the exit time (square symbols) is very close to the
proxy-breakup rate based on the diving time, that in
Fig.~\ref{fig:breakup22_63} is shown by both its discrete measurement
(star symbols) and its analytical extension provided by the Loginov's
formula (line). To understand this similarity we analyze the PDFs of
the exit time and of the diving time.
Figure~\ref{fig:exitdiving_fam63}a) shows the PDF of the diving time
for the same particle family as shown in Fig.~\ref{fig:breakup22_63}a)
and for two threshold values.  We observe that the PDF has a sharp
decrease for small diving times followed by a well developed
exponential tail.  The presence of the exponential tail implies that
the realization of diving events whose duration exceeds the
correlation time of the hydrodynamic stress can be described as a
Poisson process, where individual diving events are independent of
each other. Moreover, an exactly exponentially distributed diving time
would lead to an equivalence of the mean diving time and the mean exit
time, i.e. $\langle T\rangle=\langle \tau \rangle$ (this follows from
using $p(T)\sim {e}^{-t/\langle T\rangle}$ in Eq.~\eqref{eq:T1_T2}).
However, the deviations from the exponential distribution for short
diving events, that are particularly pronounced for the intermediate
values of the threshold shown in Fig.~\ref{fig:exitdiving_fam63},
result in the diving time being different from the exit time. These
deviations for short diving times reflect the correlation of
short-lived turbulent fluctuations, i.e. the burst in turbulent
intensity which often come with multiple spikes in the stress
intensity, as visible in Fig.~\ref{fig:signals}.  To quantify the
deviation from an exponential distribution we report in the inset of
Fig.~\ref{fig:exitdiving_fam63}a) the moment ratio $\langle
T^2\rangle/(2\langle T\rangle)$. For an exponential distribution, this
moment ratio has a value of unity. As it can be seen, the moment ratio
for the diving time is larger than one.  This has indeed to be
ascribed to the sharp drop of the diving time PDF for short diving
times.

Returning to Fig.~\ref{fig:breakup22_63}, we note that for large
thresholds values, for which breakup is governed by the rare and
intense turbulent events, there are not enough statistics to measure
the mean exit and mean diving time with confidence. In other words,
the length of exit and diving events becomes comparable or larger than
the run-time of our simulation. This causes the data for the breakup
rate based on the exit time (solid squares) and on the diving time
(star symbols) to level off or even to increase.  Inspection of the
underlying PDFs reveals that this is an artefact caused by the finite
run time of our simulation. In comparison, the Loginov's formula
(solid line) keeps on decaying, thus giving the more realistic
picture.

To independently estimate the breakup rate for large threshold values
(and thus to validate the Loginov formula) we make use of another
routine. This consist of releasing a large number of aggregates at the
beginning of the simulation and monitoring the decay of their number
concentration as breakup events occur.  From the temporal evolution of
the number of aggregates $N(t)$ we can estimate the breakup rate
assuming a first-order process as:
\begin{align}\label{eq:dNdt}
\frac{dN}{dt}=-\hat{f}_{\sigma_{cr}} N
\end{align}
On the basis of Eq.~\eqref{eq:dNdt}, the proxy breakup rate can be
determined from the slope of a plot of $\ln(N/N_0)$ versus $t$.  The
first order process given by Eq.~\eqref{eq:dNdt} assumes that the
events that cause breakup (i.e. the intense fluctuations in the
hydrodynamic stress) are independent from each other. Moreover, the
dynamics underlying Eq.~\eqref{eq:dNdt} is equivalent to an exit time
that has an exponential distribution. This is explored in
Fig.~\ref{fig:exitdiving_fam63}b) that shows the PDF of the exit time
for two threshold values. It is seen that the exit time is
approximately exponentially distributed, with small deviations at
short exit times, and at small threshold values for large exit
times. To assess the deviations from exponential distribution in the
inset of Fig.~\ref{fig:exitdiving_fam63}b) we show the normalized
second order moment of the exit time, i.e
$\langle\tau^2\rangle/(2\langle\tau\rangle)$, as a function of the
threshold stress. For an exponential distribution, the normalized
second order moment has a value of unity. As it can be seen, at
increasing threshold stress, the normalized second order moment
approaches one, implying that the exit time PDF comes closer to an
exponential distributions and hence, the dynamics proposed by
Eq.~\eqref{eq:dNdt} become more accurate.  This latter observation is
also confirmed by the breakup rate in Fig.~\ref{fig:breakup22_63}. The
open triangle symbols show the estimate based on the
Eq.~\eqref{eq:dNdt}, from which see that, for intermediate threshold
values, for which both the exit time measurement and the concentration
decay can be measured with confidence, we find good agreement between
the two. Moreover, for large threshold values the estimate based on
the concentration decay (open triangles) is in agreement with the
Loginov's formula (lines), a result which is in line with
Eq.~\eqref{eq:largethreshold}.  Based on these findings, in the
following, we will report the breakup rate in terms of the exact
expression based on the exit time for small and intermediate
thresholds, that is for breakup rates $f_{\sigma_{cr}}\tau_\eta>
0.01$, whereas for large thresholds we will report the estimate from
the decaying particle concentration of Eq.~\ref{eq:dNdt}.

\subsection{Breakup rate of inertial aggregates}
\begin{figure}\centering
\includegraphics[width=0.6\textwidth]{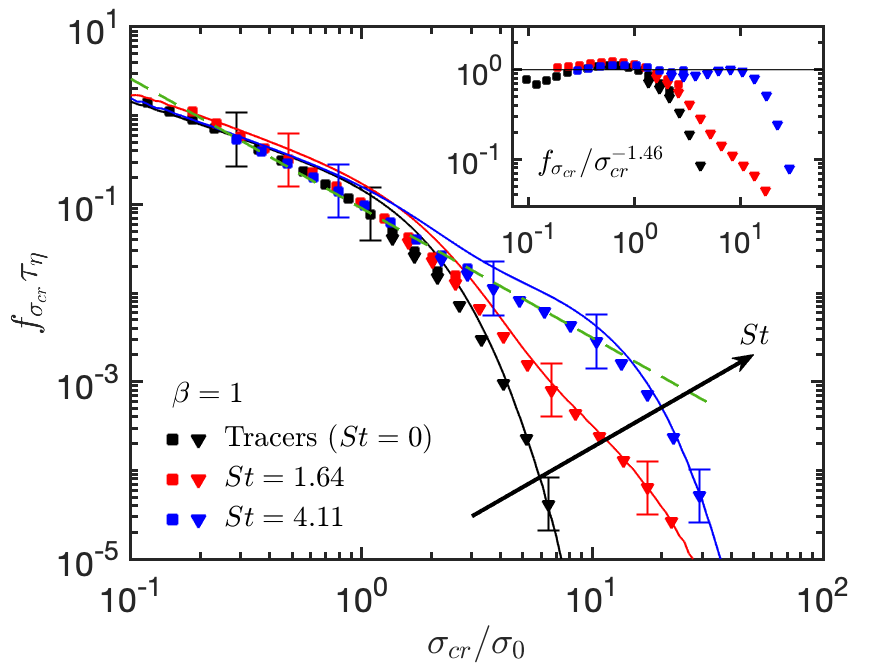}
\caption{\label{fig:breakup_beta1} 
  Breakup rate as a function of the threshold stress for $\beta=1$ and
  three different Stokes numbers measured by the exact expression
  based on the exit time (square symbols) and on the broken cluster
  approach (triangles). Solid lines report the breakup rate measured
  by the Loginov's formula. The dashed line shows the fitted power-law
  expression $f_{\sigma_{cr}}\sim \sigma_{cr}^{-k}$, with $k=1.46\pm
  0.04$.  The inset shows the breakup rate compensated by the
  power-law expression.  }
\end{figure}
Figure~\ref{fig:breakup_beta1} shows the breakup rate as a function of
the threshold stress for neutrally buoyant aggregates with varying
Stokes number. Lines refer to the proxy-breakup rate based on the
Loginov's formula of Eq.~\eqref{eq:f_loginov}, whereas symbols refer
to direct measurements based on the exit time (Eq.~\eqref{eq:f_exact};
square symbols) and on the concentration decay (Eq.~\eqref{eq:dNdt};
triangle symbols).  The data shows a clear shift in the breakup rate,
i.e., we observe a transition from a shear dominated breakup mechanism
at $St=0$ to a drag dominated one as $St$ becomes large.  Shear
stresses are relatively weak and, accordingly, only weak aggregates
are broken down by shear. This is the reason why an early drop-off of
the breakup rate occurs for tracer-like aggregates. For instance, this
makes a tracer-like aggregate of strength $\sigma_{cr}\sim
10\,\sigma_0$ very hard to break.

As the Stokes number increases, the slip velocity grows and gives
origin to an additional drag stress acting on the aggregate. This
stress causes a shift of the drop-off of the breakup rate towards
larger threshold values, such that in this case an inertial aggregate
with strength $10\,\sigma_0$ has a substantial breakup frequency, as
made apparent by the two non-tracer curves in
Fig.~\ref{fig:breakup_beta1}.

\begin{figure}
\centering
\includegraphics[width=0.5\textwidth]{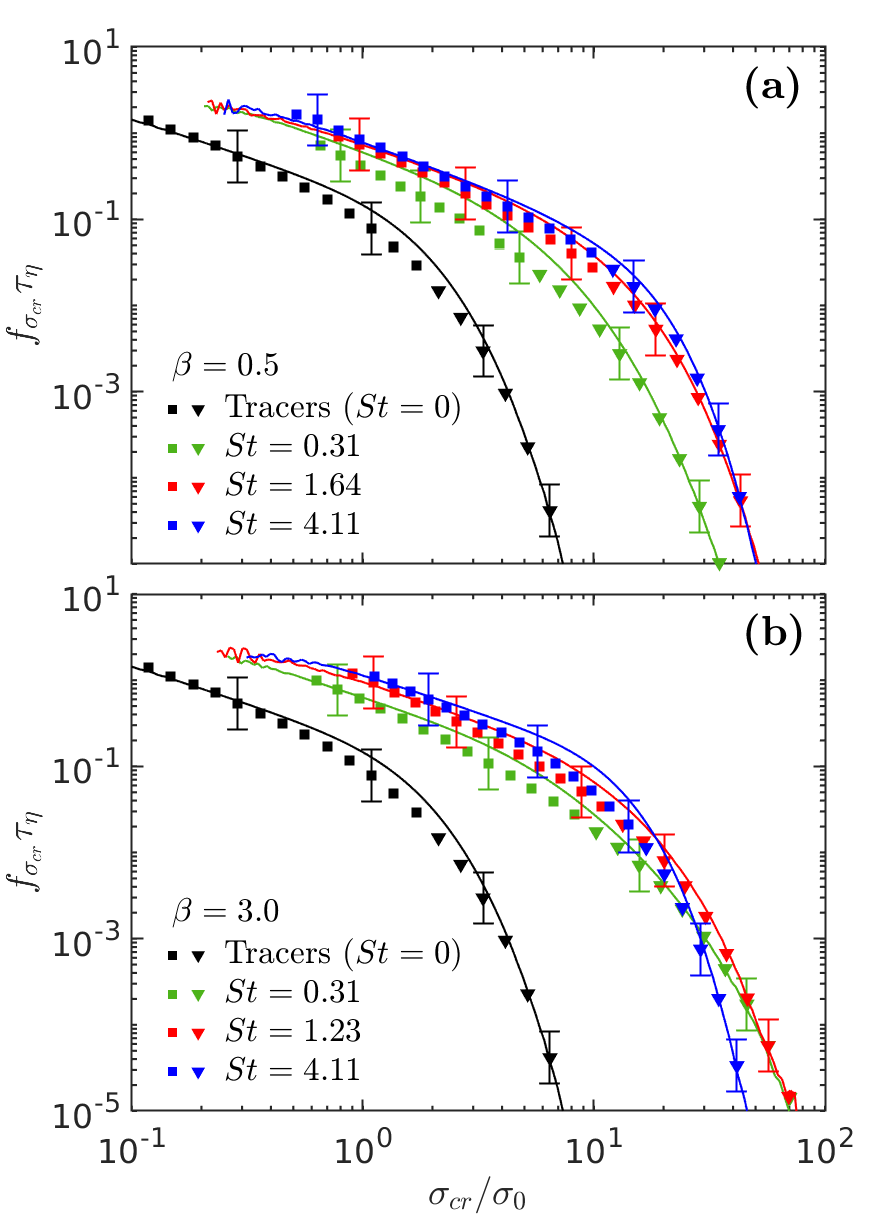}
\caption{Breakup rate for varying particle critical stress. Data refer to heavy aggregates with $\beta=0.5$ (a) and light aggregates with $\beta=3.0$ (b).}
\label{fig:loginovlight}
\end{figure}

It is interesting to notice that despite the additional contribution
of the drag stress, the power-law part of the breakup rate is not
affected by inertia, i.e., for small threshold stresses, the breakup
rates for neutrally buoyant aggregates collapse to a single master
curve following a power-law of the form $f_{\sigma_{cr}}/\tau_\eta =
0.090 \, (\sigma_{cr}/\sigma_0)^{-k}$, where $k=1.46\pm 0.04$, as
shown by the dashed curve in Fig.~\ref{fig:breakup_beta1}. The inset
of Fig.~\ref{fig:breakup_beta1} shows the breakup rate compensated by
this power-law expression, underlining the quality of the power-law
master curve.

The breakup rate for heavy and light aggregates is shown in
Fig.~\ref{fig:loginovlight}. For these aggregates, the drag stress is
dominant and causes a shift of the breakup rate towards higher
threshold stresses at already small Stokes number. For instance, at
the smallest Stokes number reported ($St=0.31$) drag stresses already
exceed the shear stresses by an order of magnitude, thus determining
the observed shift of the breakup rate towards the right. As the
Stokes number further increases, the breakup rate for both heavy and
light aggregates saturates. For heavy particles ($\beta=0.5$,
Fig.~\ref{fig:loginovlight}a)) the breakup rate reaches an apparent
maximum. This maximum (made apparent by the overlap of the curves for
$St=1.64$ and $St=4.11$ in Fig.~\ref{fig:loginovlight}a) follows again
a power-law for the small threshold values and has a sharp
super-exponential cut-off at larger thresholds.  For light particles
($\beta=3$, Fig.~\ref{fig:loginovlight})) the overlap at large Stokes
numbers is less pronounced and we even observe a reduction in the
breakup rate for large threshold stresses and large Stokes
numbers. This sharper drop-off at $St=4.11$ well agrees with the PDFs
of the total stress reported in Fig.~\ref{fig:PDF}f).  Careful
inspection of the PDFs reveals in fact a slightly wider right tail at
intermediate Stokes numbers (i.e. $St=1.64$) compared to the largest
Stokes number studied in our work ($St=4.11$). This implies that large
excursions of the total stress are more frequent for intermediate
Stokes numbers than for large Stokes numbers, i.e., light particles
with large Stokes number exhibit a mild filtering effect that prevents
them from experiencing high drag stresses, thus causing the sharper
cut-off of the breakup rate for $St=4.11$ visible in
Fig.~\ref{fig:loginovlight}b.

\section {Conclusions}
\label{sec:closeup}
In this work we have studied the stress and breakup statistics of
inertial aggregates in homogeneous isotropic turbulence, at varying
fluid-to-particle density ratio and Stokes number. We have solved the
flow dynamics by a direct numerical simulation and tracked the
particles trajectories by evolving a minimal formulation of the
Maxey-Riley-Gatignol equation.  We have deemed both shear stress and
drag stress as responsible for the particle breakup and we have
assumed breakup to occur in a brittle manner, i.e., aggregates break
instantaneously when they experience a fluid dynamic stress larger
than their mechanical resistance. We have devised and tested different
approaches for measuring the breakup frequency, discussing in detail
their theoretical foundations and limitations.

Two distinct breakup regimes exist, depending on the aggregate
mechanical strength. Loose aggregates have large breakup frequencies
and are broken down in the smooth regions of the flow, where the
stresses are Gaussian distributed, thus making the breakup frequency
to follow a power-law behaviour with the aggregate
strength. Conversely, strong aggregates have lower breakup rates and
are broken down by the burst of the hydrodynamic stresses, which are
dictated by the turbulent intermittency.
 
Results have also shown that inertial effects have a major role in
determining breakup rates. When inertial effects are limited (i.e.,
for neutrally buoyant particles with small Stokes number), particles
behave very similarly to tracers: they similarly sample the flow
topology and their breakup is mostly dictated by the shear stress
statistics. However, as soon as deviations from neutral buoyancy
occur, inertial effects become dominant even at small Stokes numbers
and have been seen to be able to cause the breakup of even the most
resistant aggregates.  A characterization of the flow topology seen by
particles has also been conducted, and it agrees quantitatively with
previously reported data for turbulence induced segregation.

Our investigation puts the basis for further developments for the
measure and modeling of breakup due to shear and drag stresses. In
fact, it can be considered as a first step towards the development of
breakup kernels for inertial brittle aggregates. Laboratory and
numerical investigations aimed at assessing the outcome of breakup
events in terms of the fragment size distribution would complement our
findings, and provide the full information needed for calibrating
macroscopic population balance models addressing breakup in
homogeneous solid-liquid turbulent flows.

\section*{Acknowledgments}
 We thank Marco Vanni for useful discussions and
 comments. L. B. received funding from the European Research Council
 (ERC) under the European Union’s Horizon 2020 research and innovation
 programme (grant agreement No 882340). G. F. gratefully acknowledges
 support from the “TUM Global Postdoc Fellowship”. Numerical
 simulations were performed at CINECA (Italy).

\end{document}